\documentclass[amsmath,twocolumn,nofootinbib]{revtex4}
\usepackage[pdftex]{graphicx}

\begin{document}
 
\title{Bayesian Limits on Primordial Isotropy Breaking}
\author{C. Armendariz-Picon and Larne Pekowsky}
\affiliation{Physics Department, Syracuse University, \\
Syracuse NY 13244-1130, USA}

\begin{abstract}

It is often assumed that primordial perturbations are statistically isotropic, which implies, among other properties, that their power spectrum is invariant under rotations.  In this article, we test this assumption by placing model-independent bounds on deviations from rotational invariance of the primordial spectrum. Using five-year WMAP cosmic microwave anisotropy maps, we set limits on the overall norm and the amplitude of individual components of the  primordial spectrum quadrupole.  We find that there is no significant evidence for primordial isotropy breaking, and that an eventually non-vanishing quadrupole  has to be subdominant.

\end{abstract}
\maketitle

\section{Introduction}

Observations of the cosmic microwave background show that the early universe contained tiny density perturbations, from which structures developed as the universe expanded.  In recent times, we have gained a wealth of information about  these primordial perturbations. We have precisely measured their spectrum, and we have placed quite stringent limits on their properties \cite{Komatsu:2008hk}. Prompted by these advances, the nature of the primordial perturbations has entered the standard cosmological model, a set of a few parameters and assumptions that summarizes what we know about our universe.

There is however an assumption in the standard cosmological model that has not been subject to much observational or theoretical scrutiny: the statistical isotropy of the primordial perturbations. Cosmological perturbations are statistically isotropic if their probability distribution functionals are invariant under rotations, which implies in particular that the power spectrum of statistically isotropic perturbations only depends on the magnitude of the wave vector, $\mathcal{P}(\mathbf{k})=\mathcal{P}(k)$. Though some papers have analyzed the impact of statistically anisotropic perturbations on structure \cite{ArmendarizPicon:2005jh,Pullen:2007tu, Ando:2008zz}, while others have proposed mechanisms for their generation \cite{Ackerman:2007nb,ArmendarizPicon:2007nr,Pereira:2007yy,Gumrukcuoglu:2007bx,Akofor:2007fv,Erickcek:2008sm,Yokoyama:2008xw},  to date no model-independent limits on the deviations from statistical isotropy of the primordial perturbations exist.\footnote{On the other hand, the statistical isotropy of the CMB itself has been extensively investigated: see \cite{Souradeep:2006dz} and references therein.}  In this article we set  precise bounds, and thus verify one of the key ingredients in our understanding of the origin of structure.

\section{Statistical Anisotropy}

In order to study deviations from statistical isotropy, we need to find an appropriate way to parametrize those deviations. Following \cite{ArmendarizPicon:2005jh}, we expand the primordial power spectrum $\mathcal{P}_\mathcal{R}(\mathbf{k})$ in spherical harmonics,\footnote{We adhere to  the normalization conventions of \cite{Komatsu:2008hk}, which differ from those of \cite{ArmendarizPicon:2005jh}.}
\begin{equation}\label{eq:P}
	\mathcal{P}_\mathcal{R}(\mathbf{k})=\sqrt{4\pi}\sum_{\ell m} \mathcal{P}_{\ell m}(k)Y_{\ell m}(\hat{k}),
\end{equation}
which in fact is the most general form the  spectrum can take. For statistically isotropic perturbations only the ``monopole" $\mathcal{P}_{00}$ is non-zero, whereas a non-vanishing $\mathcal{P}_{\ell m}$ for $\ell\neq 0$ is what characterizes statistically anisotropic Gaussian perturbations. 

We assume that the multipole components can be approximated by power laws,
\begin{equation}
	\mathcal{P}_{\ell m}(k)=\mathcal{A}_{lm}\cdot \left(\frac{k}{2 \cdot 10^{-3} \text{Mpc}^{-1}}\right)^{n_s-1},
\end{equation}
with a common spectral index $n_s$. This is in fact what  many of the models of primordial isotropy breaking predict \cite{Ackerman:2007nb,ArmendarizPicon:2007nr,Yokoyama:2008xw}. In any case, because the range of scales we consider is relatively small, our results should also apply for mildly scale-dependent primordial spectrum multipoles, even if they do not share the same spectral index.
Note that the $\mathcal{A}_{\ell m}$ are not completely arbitrary:  by definition the power has to be positive definite and invariant under spatial inversion. The first condition requires that the $\mathcal{A}_{\ell m}$ be a ``square", while last condition implies the vanishing of  $\mathcal{A}_{\ell m}$ for odd values of $\ell$.   For our purposes, however, it  suffices to treat the $\mathcal{A}_{\ell m}$ (for even $\ell$) as free parameters.

We shall use cosmic microwave measurements to put constraints on the multipoles of the power spectrum. The reader should be aware that the multipole space we have been considering here is quite different from the multipole space of the  temperature anisotropies. Nevertheless, the two are not completely independent. A statistically anisotropic power spectrum  induces correlations between temperature multipoles with different values of $\ell$  \cite{ArmendarizPicon:2005jh},
\begin{multline} \label{eq:A}
	\langle a^*_{\ell_1 m_1} a_{\ell_2 m_2}\rangle = 4\pi(-i)^{\ell_2-\ell_1}\sum_{\ell m} D(\ell_1 m_1; \ell m;  \ell_2 m_2) \times \\
	\times \int \frac{dk}{k} \Delta_{\ell_1} \Delta_{\ell_2} \mathcal{P}_{\ell m}(k),
\end{multline}
where the $\Delta_\ell$ are the radiation transfer functions, and $D$ is a product of Clebsch-Gordan coefficients. Note that equation (\ref{eq:A}) also applies for multipole expansions in  real spherical harmonics, \begin{equation}
Y^\text{real}_{\ell m}\equiv \bigg\{
	\begin{array}{lc}
		\sqrt{2}\, \text{Re}\, Y_{\ell m}, & m \geq 0 \\ 
		\sqrt{2}\, \text{Im}\,  Y_{\ell -m}, &  m < 0.
	\end{array}
\end{equation}
For convenience we shall work here with the latter. In that case both  $a_{\ell m}$ and $\mathcal{P}_{\ell m}$ are real, and $D$ is a linear combination of the complex $D$, determined by the unitary transformation that relates real and complex spherical harmonics.

\section{Bayesian Analysis}

In this work we follow a Bayesian approach to inference, that is, we consider the posterior probability of the amplitudes $\mathcal{A}_{\ell m}$, given that we observe the temperature anisotropies $\mathbf{a}$,
\begin{equation}\label{eq:posterior}
	P(\mathcal{A}_{\ell m}|\mathbf{a})\propto L(\mathbf{a}| \mathcal{A}_{\ell m})\, P(\mathcal{A}_{\ell m}).
\end{equation}
The function $L(\mathbf{a}| \mathcal{A}_{\ell m})$ is the likelihood, and $P(\mathcal{A}_{\ell m})$ is the prior. For notational convenience we gather the pair of multipole indices $(\ell, m)$ into a single index $\alpha$, and we collect all the temperature anisotropies in a vector $\mathbf{a}$, with components $a_\alpha$.

Unfortunately, full-sky maps of the cosmic microwave background with well-defined error properties do not exist, because galactic contamination cannot be reliably removed from some regions of the sky. We are thus forced to deal with masked skies $\mathbf{c}$, from  which those regions are excluded,
\begin{equation}\label{eq:real}
	c(\hat{r})= M(\hat{r}) \cdot \frac{\delta T}{T}(\hat{r}).
\end{equation}
The function  $M$ is the mask and $\delta T/T$ are the temperature anisotropies. To proceed further, it is useful to have the  counterpart of equation (\ref{eq:real}) in multipole space, which can be readily shown to be
\begin{multline}\label{eq:c}
	\mathbf{c}= M \, \mathbf{b}, 
	\quad \text{where} \\
	 M_{\ell m,\ell_1 m_1}=\sum_{\ell_2 m_2} \frac{D(\ell m; \ell_1 m_1; \ell_2 m_2)}{\sqrt{4\pi}} \,M_{\ell_2 m_2},
\end{multline} 
and the $b_{\ell m}$ are the spherical harmonic coefficients of the unmasked sky map.

But our troubles do not end here. In a real experiment, instruments have noise, and beams do not have infinite resolution. In addition, temperature maps are not provided as smooth functions over the sky, but rather, as pixelized functions over the sphere.  Adding the instrument noise  $n$ to the cosmic microwave temperature, $T_{\text{observed}}=T_{\text{CMB}}+n$, and convolving the signal with, respectively,  the instrument and pixel window functions $W$ and $H$ we arrive at the multipoles of the temperature map,
\begin{equation}\label{eq:b}
	\mathbf{b}=H\, W \,\mathbf{a}+\mathbf{n}.
\end{equation}
We assume that both $W$ and $H$ are diagonal and $m$-independent. In particular, we do not take beam  and pixel asymmetries into account.

The only problem left is to calculate how likely a particular temperature vector $\mathbf{c}$ is. Assuming that the temperature multipoles are Gaussian, inserting equation (\ref{eq:b}) into (\ref{eq:c}), and substituting into the analogue of equation (\ref{eq:posterior}) we obtain 
\begin{equation}\label{eq:Pc}
	P(\mathcal{A}_{\ell m}|\mathbf{c})\propto \frac{1}{\det{}^{1/2} C}\exp\left(-\frac{1}{2}\mathbf{c} \cdot C^{-1} \mathbf{c}\,\right) \, P(\mathcal{A}_{\ell m}),
\end{equation}
where $C$ is the covariance matrix of the masked temperature multipoles,
\begin{equation}
	C= (M H W)  A  (M H W)^T + M N M^T,
\end{equation}
and $A$ is the covariance matrix of the unmasked temperature anisotropies, equation (\ref{eq:A}).  The matrix $N$ is the pixel noise covariance matrix, with multipole components
\begin{equation}
	N_{\ell_1 m_1, \ell_2 m_2}= \Delta a \sum_{\ell m} \frac{D(\ell_1 m_1; \ell_2 m_2; \ell m)}{\sqrt{4\pi}} N_{\ell m},
\end{equation}
where $\Delta a$ is the area of each pixel in the temperature map, and $N_{\ell m}$ is the discrete spherical harmonic transform of the noise variance,\begin{equation}
	N_{\ell m}=\sum_{i} \Delta a\, N_i \,  Y_{\ell m}(\hat{r}_i),
	\quad  \text{where} \;\; \langle n_i n_j \rangle\equiv N_i \delta_{i j}.
\end{equation}
The indices $i$ and $j$ run over all the pixels on  the sphere.

\section{Data and Implementation}

\subsection{Data}

We analyze the five-year WMAP  foreground-reduced V2 and W1 differential assembly temperature maps \cite{Komatsu:2008hk}. These are the maps with the lowest noise in the V and W frequency  bands, which are the ones less exposed to foreground contamination. We do not consider combined frequency band maps here because they are averages of individual differential assemblies with direction-dependent weights.  In general, for such averages the matrix $W$ is not diagonal.

Because of computational limitations, it is not possible to analyze all the data in the maps. If we restrict our analysis to temperature multipoles with $\ell\leq \ell_\text{max}$ we need to consider covariance matrices with a number of elements that scales like $\ell_\text{max}^4$, which quickly become intractable as $\ell_\text{max}$ grows. In reality, one has to work with even larger matrices, because the masking aliases power from high multipoles to low multipoles. If the mask is band-limited at $\ell^M_\text{max}$, by equation (\ref{eq:b}) the multipole $\ell_\text{max}$ of the masked  sky contains contributions from unmasked multipoles at $\ell_\text{max}+\ell^M_\text{max}$. Hence, not only do we have to limit the amount of masked multipoles, but also the mask bandwidth. We restrict our analysis to masked temperate multipoles up to $\ell_\text{max}=62$, and to a mask with bandwidth $\ell^M_\text{max}=92$.  Some of the masked multipoles have to be discarded however, as described below.

The noise power in the WMAP temperature maps is anisotropic, because different parts of the sky are observed a different number of times. Though  the monopole $\ell=0$ is the dominant noise component, we also keep multipoles with $\ell=2$, to make sure that anisotropies in the noise do not creep into our estimate of the primordial spectrum quadrupole.  In any case, at $\ell_\text{max}+\ell^M_\text{max}=154$, our errors are dominated by cosmic variance. 

\subsection{Mask}
Starting from the WMAP 5-year temperature analysis/KQ85 mask\footnote{This mask is available at {\tt http://lambda.gsfc.nasa.gov}} at HEALPix\footnote{The HEALPix web site is at {\tt http://healpix.jpl.nasa.gov}} resolution ${N_\text{side}=512}$, we construct our analysis mask by sequentially following these steps:
$i)$~Smooth with a Gaussian beam of {FWHM=$1440$} arcmin,
$ii)$~set pixels $i$ with  ${M_i<0.92}$ to 0, and to 1 otherwise,
$iii)$~smooth again with a Gaussian beam of {FHWM=$492$} arcmin, and, finally, $iv)$~set mask multipoles with ${\ell>\ell^M_\text{max}=92}$ to zero. 
The first step degrades the mask and eliminates its small scale features. The second ensures that the degraded mask still masks the galactic region. The third step removes the substructure introduced by  step two. And the last step ensures that the mask is band-limited at the desired multipole value.

\begin{figure}
	\includegraphics[width=0.3\textwidth]{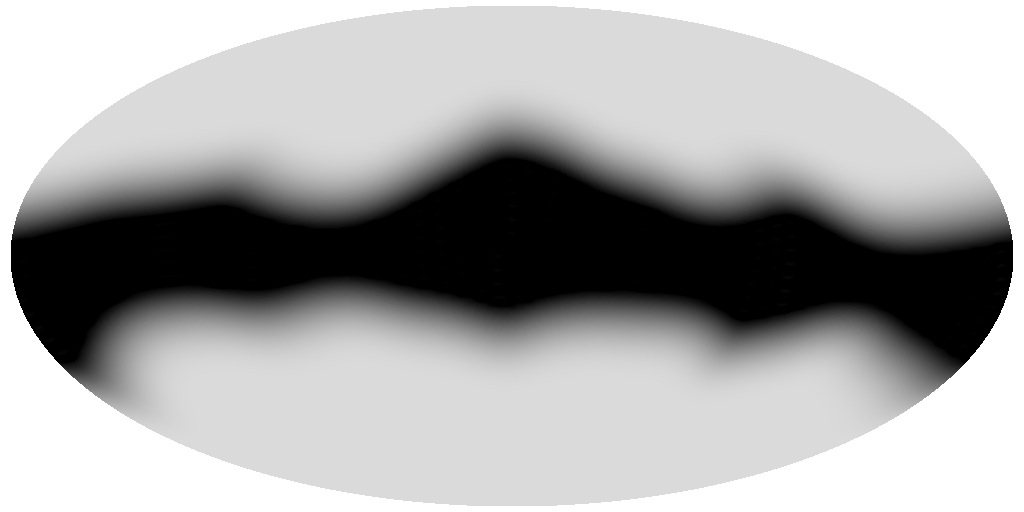}
	\includegraphics[width=0.3\textwidth]{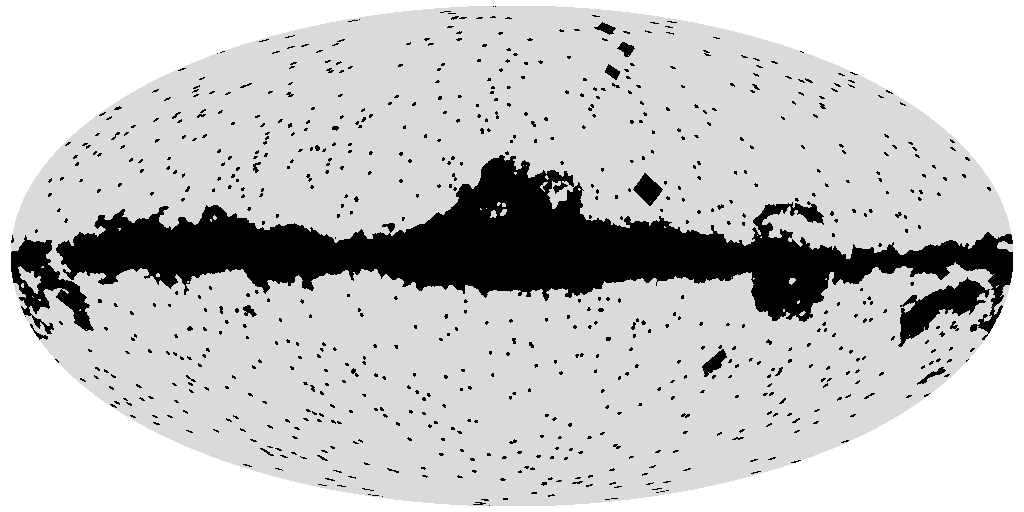}
	\caption{Logarithm of the absolute value of our analysis mask (top) and actual KQ85 mask (bottom). The absolute value of our mask in the innermost (black) regions of the galaxy is smaller than $10^{-9}$. Our analysis mask covers $55\%$ of the sky. \label{fig:mask}}
\end{figure}

Figure \ref{fig:mask} shows the logarithm of the absolute value of our mask. Because our mask is band-limited, it cannot reproduce the the original KQ85 mask. In particular, it does not cover the catalogued point sources, and it does not exactly vanish in the contaminated galactic region. To quantify the bias caused by an eventual galactic or point source contamination, we simulate 25 statistically isotropic random maps, and set the temperature of those pixels that would have been excluded by the original KQ85 mask to its value in the actual foreground-reduced V2 map. We then mask this artificial maps with our degraded mask and estimate the values of the amplitudes $\mathcal{A}_{\ell m}$ using our analysis pipeline. Their weighted means are collected in the last column of Table \ref{tab:limits}. Of course, it is still possible for unresolved point sources to further contaminate our data, but this contamination is expected to be small at our resolution \cite{Pierpaoli:2003yy}. To make sure that our bounds do not depend on the mask, we  repeat our analysis using a  mask with $\ell^M_\text{max}=98$ and $60\%$  sky coverage, and verify that this change does not significantly alter our results.

\subsection{Markov Chain Monte Carlo}

We sample the posterior probabilities in equation (\ref{eq:Pc}) with Monte Carlo Markov chains of $4\cdot 10^4$ elements. Thus, the $95\%$ credible intervals we quote in the next section actually correspond to about $95\pm 0.1\%$ probability content.  We check for convergence of our chains using the spectral analysis method described in \cite{Dunkley:2004sv}. All our chains satisfy the convergence criteria described therein. We pick the starting point from previous runs, so no burn-in period is needed

Because the matrix $M$ is ill-conditioned, one cannot accurately calculate the inverse of $C$ in the  likelihood function (\ref{eq:Pc}) numerically. Instead, we determine the singular value decomposition of $M^T$, $M^T=U\cdot \Sigma \cdot V^T$, and consider the likelihood with $\mathbf{d}= V^T \mathbf{c}$ as data. Those modes $d_\alpha$ with $\Sigma_{\alpha\alpha}< \Sigma_{11}\cdot\sqrt{c(A)\cdot f}$ are removed from the analysis. The factor $f=2.2\cdot 10^{-16}$ is the floating number precision of our computer, and $c(A)\approx 10^3$ is the condition of the matrix $A$. The cut keeps 3523 out of 3965 modes. 

We calculate the radiation transfer functions in equation (\ref{eq:A}) with a modified version of CMBEASY \cite{Doran:2003sy}. Since we fix the cosmological model, the transfer functions have to be computed only once. 
 
\section{Results}

\begin{figure}
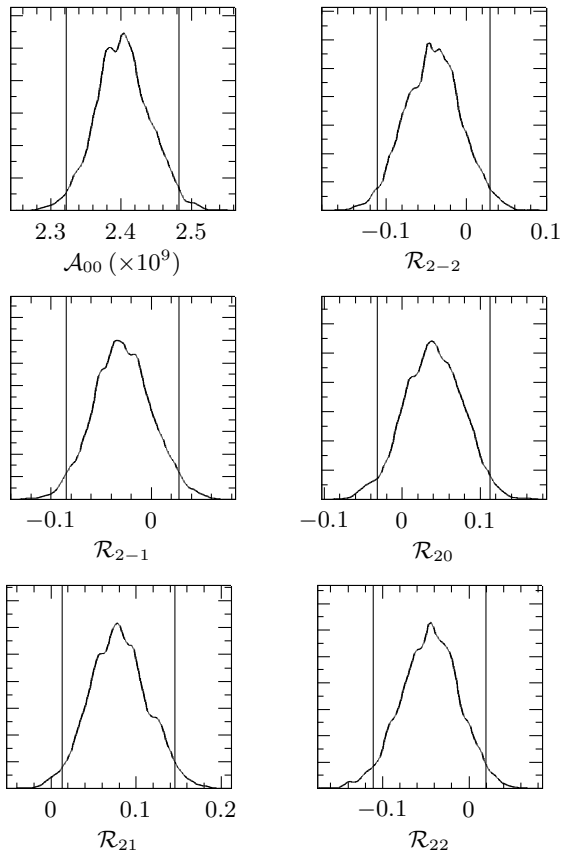

	\input{hist0.tex}
	\input{hist1.tex}
	\input{hist2.tex}
	\input{hist3.tex}
	\input{hist4.tex}
	\input{hist5.tex}  
	\caption{Kernel-smoothed marginalized posterior probability distributions. The vertical lines mark mean-centered  $95\%$ credible intervals. Note that the bias caused by galaxy and point sources has not been removed.}
	\label{fig:posteriors}
\end {figure}

Previous work on cosmological parameter estimation has focused on the properties of the scalar component of the power spectrum $\mathcal{P}_{00}$. We proceed beyond the monopole, and constrain the simplest deviation from statistical isotropy, namely, the quadrupole of the primordial spectrum $\mathcal{P}_{2 m}$.  As discussed in \cite{ArmendarizPicon:2005jh}, or just by symmetry, the expectation of scalar estimators of the temperature anisotropy multipoles $C_\ell$ only depends on  $\mathcal{P}_{00}$. Hence, we can trust the cosmological parameters derived from fits to the angular spectrum $C_\ell$ even if primordial perturbations are statistically anisotropic. Here we use the $\Lambda$CDM parameters listed in the  WMAP five-year cosmological parameter table at \cite{lambda}. To check whether our limits depend on the assumed cosmological model, we repeat the analysis with the WMAP five-year $\Lambda$CDM+TENS parameter set \cite{lambda}. Within statistical errors, our results do not change.
 
Rather than directly constraining the amplitude of the anisotropic components of the power spectrum, it is more convenient to study the posteriors of the monopole $\mathcal{A}_{00}$ and the ratios
\begin{equation}
	\mathcal{R}_{2 m}\equiv \frac{\mathcal{A}_{2 m}}{\mathcal{A}_{00}}.
\end{equation}
In the analysis of the V2 and W1 maps we impose a Gaussian prior on $\mathcal{A}_{00}$ based on its determination by the WMAP experiment \cite{Komatsu:2008hk}, and flat priors on the remaining parameters $\mathcal{R}_{2 m}$.  If we replace the former by a uniform prior, we obtain nearly the same limits. The agreement of our constraints on $\mathcal{A}_{00}$ with those of the WMAP team provides a reassuring consistency check.  We also analyze the W1 map using Gaussian priors derived from the results of our V2 analysis. This is  what we label as ${W1|V2}$. The  posterior distributions of the parameters in ${W1|V2}$ are plotted in Figure \ref{fig:posteriors}. Sample mean and $95\%$ credible intervals derived from the V2, W1 and ${W1|V2}$ runs are listed in Table \ref{tab:limits}, along with the bias caused by the imperfect mask.  In order to quantify the overall magnitude of the quadrupole, we use its norm
\begin{equation}\label{eq:A2}
	||\mathcal{R}_2||\equiv \sqrt{\, \sum_m \mathcal{R}^2_{2m}},
\end{equation}
which is invariant under rotations. 

\begin{table}
$
\begin{array}[t]{ c r@{\,\pm\,}l r@{\,\pm\,}l r@{\,\pm\,}l r@{.}l }
 \hline \hline
\text{Parameter} & \multicolumn{2}{c}{\text{V2}} & 
\multicolumn{2}{c}{\text{W1}} & 
\multicolumn{2}{c}{W1 \,|\,  V2}  & \multicolumn{2}{c}{\text{Bias}}\\
 \hline \hline
 \mathcal{A}_{00}\times 10^9 & 2.42 &  0.11  &  2.39 & 0.11 &  2.40  & 0.08 & 0&02\\
 \mathcal{R}_{2-2} & -0.04 & 0.09 & -0.03 & 0.10 & -0.04 & 0.07  & 0&01\\
 \mathcal{R}_{2-1} & -0.03 & 0.08 &  -0.02 & 0.08 & -0.03 & 0.06 & -0&03 \\
 \mathcal{R}_{20} & 0.05  & 0.10 & 0.04 & 0.10 & 0.04 & 0.07 & -0&02 \\
 \mathcal{R}_{21} & 0.08 & 0.09 & 0.08&  0.09 & 0.08 & 0.07 & 0&03 \\
 \mathcal{R}_{22} & -0.07 & 0.10 & -0.02 & 0.09 & -0.04 & 0.07 & -0&05 \\
 ||\mathcal{R}_2||  & \multicolumn{2}{c}{<0.24} &\multicolumn{2}{c}{<0.22}
 			& \multicolumn{2}{c}{<0.19} \\
 \hline\hline
\end{array}
$
\caption{Sample mean and  $95\%$ credible intervals from the posterior distributions. In the last column we also list an estimate of the bias caused by galactic and point source contamination.}
\label{tab:limits}
\end{table}

We might also extend our analysis to assess whether statistically isotropic perturbations are a better model for the data. Since we cannot compute the Bayesian evidence within our approach, we determine instead three non-exclusively Bayesian measures that have been widely used in the literature: the effective chi squared,  ${\chi^2_\text{eff}\equiv -2 \log L_\text{max}}$, the Akaike Information Criterion (AIC) and the Bayes Information Criterion (BIC) (see for instance \cite{Liddle:2007fy}.) Their differences under the assumptions of a non-vanishing and vanishing quadrupole are listed in Table \ref{tab:criteria}.

\begin{table}
$
\begin{array}[t]{ l r@{.}l  r@{.}l }
 \hline \hline
\text{Criterion} & \multicolumn{2}{c}{\text{V2}} & \multicolumn{2}{c}{\text{W1}}  \\
 \hline \hline
 \Delta\chi^2_\text{eff} & -6&0 & -3&2 \\
 \Delta \text{AIC} &4&0 & 6&8\\
 \Delta \text{BIC} & 34&8 &  37&6 \\
 \hline\hline
\end{array}
$
\caption{Comparison between fits to the data with a non-vanishing and a vanishing quadrupole, $\Delta X= X_\text{ani}- X_\text{iso}$. The inclusion of a quadrupole in the primordial power spectrum requires five additional parameters.}
\label{tab:criteria}
\end{table}

\section{Conclusions}
Inspection of Table \ref{tab:limits} quickly reveals that the amplitude of the quadrupole components  is consistent with statistical isotropy. In particular, if there is a non-vanishing quadrupole in the primordial spectrum, it clearly has to be subdominant. The results in Table \ref{tab:criteria} also imply that there is no evidence for primordial statistical anisotropy. Although a non-zero primordial quadrupole significantly increases the likelihood, the information criteria that penalize the introduction of additional parameters strongly favor isotropy.

Because any deviation from statistical isotropy can be cast as in equation (\ref{eq:P}), the limits that we have found are mode-independent, and can thus be directly applied to any of the models for the generation of (adiabatic) statistically anisotropic perturbations discussed in the literature \cite{Ackerman:2007nb,ArmendarizPicon:2007nr,Gumrukcuoglu:2007bx,Akofor:2007fv,Erickcek:2008sm,Yokoyama:2008xw}. We have not studied how our bounds constrain the parameters of these models, but it should be straight-forward to do so. On the other hand, our null results confirm again the predictions of the simplest inflationary models.  

The study of the statistical isotropy of the primordial perturbations is still in its infancy, and our analysis is just a first step toward  preciser measurements of the primordial spectrum.  With more data, improved analysis techniques, and better control of systematics, it should be possible in principle to obtain much tighter constraints \cite{Pullen:2007tu}.

\section*{Note added}
Shortly before submission of this manuscript, a preprint with significant overlap with the work presented here appeared on the arXiv \cite{Groeneboom:2008fz}.

\begin{acknowledgments}

We acknowledge the use of the Legacy Archive for Microwave Background Data Analysis (LAMBDA). Support for LAMBDA is provided by the NASA Office of Space Science. Some of the results in this paper have been derived using the HEALPix \cite{Gorski:2004by} package. The work of CAP is supported in part by the National Science Foundation under grant PHY-0604760.
\end{acknowledgments}

\end{document}